\documentclass[12pt,draftnote]{article}
\usepackage{fullpage}
\usepackage{graphics,xspace,verbatim,amsmath}
\usepackage[squaren]{SIunits}
\usepackage{hyperref}

\hypersetup{
    colorlinks=true,       
    linkcolor=red,          
    citecolor=blue,        
    filecolor=blue,      
    urlcolor=blue           
}

\newcommand{\ben}{\begin{equation}}
\newcommand{\een}{\end{equation}}
\newcommand{\bean}{\begin{eqnarray}}
\newcommand{\eean}{\end{eqnarray}}
\newcommand{\be}{\[}
\newcommand{\ee}{\]}
\newcommand{\bea}{\begin{eqnarray*}}
\newcommand{\eea}{\end{eqnarray*}}

\begin{document}

\title{Tutorial on agent-based models in Netlogo\\ applied to immunology and virology}

\author{%
Catherine A.\ A.\ Beauchemin (\href{mailto:cbeau@ryerson.ca}{cbeau@ryerson.ca})$^{*,\dagger}$,\\%
Laura E.\ Liao$^*$, and Kenneth Blahut$^*$\\
{\small $^*$Department of Physics, Ryerson University}\\[-0.3em]%
{\small $^\dagger$Interdisciplinary Theoretical and Mathematical Sciences (iTHEMS) Program}%
}

\date{%
\begin{tabular}{rl}
Last revised:& \today \\
NetLogo version:& 6.0
\end{tabular}}

\maketitle

\begin{abstract}
This tutorial introduces participants to the design and implementation of an agent-based model using NetLogo through one of two different projects: modelling T cell movement within a lymph node or modelling the progress of a viral infection in an in vitro cell culture monolayer. Each project is broken into a series of incremental steps of increasing complexity. Each step is described in detail and the code to type in is initially provided. However, each project has room to grow in complexity and biological realism so participants are encouraged to expand their project beyond the scope of the tutorial or to develop a project of their own.
\end{abstract}

\subsection*{Suggested reading}

In preparation for the tutorial, participants are encouraged to read:
\begin{itemize}
\item N Moreira. In pixels and in health: Computer modeling pushes the threshold of medical research. Science News, 169(3):40--44, 21 Jan, 2006
\item C Beauchemin, J Samuel, J Tuszynski. A simple cellular automaton model for influenza A viral infections. J.\ Theor.\ Biol., 232(2):223--234, 21 Jan, 2005.\\doi:\href{http://dx.doi.org/10.1016/j.jtbi.2004.08.001}{10.1016/j.jtbi.2004.08.001}
\item DM Catron, AA Itano, KA Pape, DL Mueller, and MK Jenkins. Visualizing the first 50 hr of the primary immune response to a soluble antigen. Immunity, 21(3):341--347, Sep, 2004. doi:\href{http://dx.doi.org/10.1016/j.immuni.2004.08.007}{10.1016/j.immuni.2004.08.007}
\end{itemize}
to understand the basics of agent-based models of infection and/or immunity.


\newpage
\tableofcontents

\newpage

\section{NetLogo: A basic introduction}

\begin{center}
\resizebox{0.8\linewidth}{!}{\includegraphics{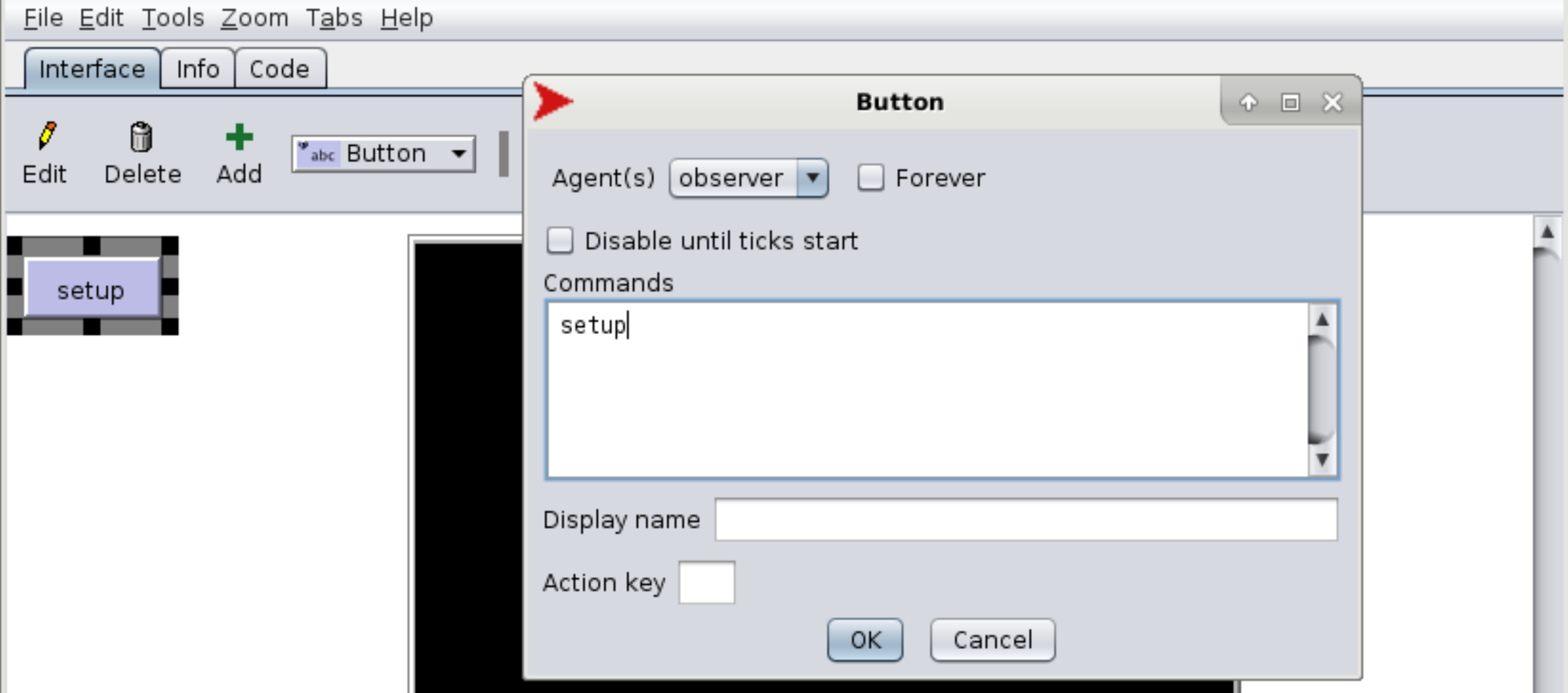}}
\end{center}

\subsection{Getting NetLogo}

To obtain NetLogo for your computer operating system (Mac, Linux, Windows), visit the NetLogo website at \url{http://ccl.northwestern.edu/netlogo}.

\subsection{Looking at a finished model from the Models Library}

We'll now have a look at what an example finished client looks like.
\begin{enumerate}
\item Select ``File'' $\rightarrow$ ``Models Library''
\item Select ``Sample Models'' $\rightarrow$ ``Biology'' $\rightarrow$ ``Tumor''
\end{enumerate}
You see 3 tabs:
\begin{description}
\item[Interface:] Where the simulation gets displayed and the end-user can interact with it using the available sliders, buttons, etc.\ which were made available by the author of the model.
\item[Info:] Where you can find general information about the model: what it is, how to use it, etc. It is essentially the help page for the model.
\item[Code:] Where the NetLogo code driving the model is located. You can modify it and see what happens or you can write brand new code.
\end{description}

\subsubsection{Exploring the Interface tab}

The `setup' button is a standard button in NetLogo and it is used to initialize the model, i.e., set up the starting conditions for your model. For example, how many cells do you want initially, where should they be, what colour should they be, what should be their initial state (e.g., infected, uninfected, dead).

The `go' button does exactly what you'd expect: it makes the simulation go. But be sure to press `setup' before you press `go'.

The remaining buttons allow you to trigger certain actions during the course of the simulations. The various sliders and switches allow you to adjust various parameters or conditions of the simulations: You can play around with them even as the simulation is running. Finally, the monitors and plots are used to display variables of the simulations (e.g., how many cells there are).

Note that if you want to restart the simulation, you can press `go' to stop the currently running simulation and then press `setup' to reinitialize it.

Now that you know how to run a model, let's learn how to make one.


\subsection{Some NetLogo basics}

The NetLogo User Manual is available at \url{http://ccl.northwestern.edu/netlogo/docs/}. Here, I will provide a quick intro to help you get started.

A NetLogo simulation consists of a world made up of rectangles (in 2D) or blocks (in 3D) called `patches' within which mobile agents called `turtles' can move and evolve based on the turtles and patches around them and those encountered along their path. The patches can also evolve; more on that later.

There are 3 types of agents of importance for us in NetLogo:
\begin{description}
\item[Patch:] A small rectangle (in 2D) or block (in 3D) of the simulation world. A patch is identified by its coordinate, $(x,y)$ or $(x,y,z)$. Thus it CANNOT move but it can hold variables (e.g., its colour, how long ago it was visited, what type of a site it is). You cannot have different breeds of patches: there is only one set of patches making up your world.
\item[Turtle:] A mobile agent which can go from one patch to another based on whatever rules are defined for its motion. Because it can move, a turtle agent is identified not by its coordinate but instead by a ``who'' number. Turtles can also hold variables, and you can have different breeds of turtles (different types of turtle agents, e.g., rabbits and hares).
\item[Link:] A connection which can be created between two turtle agents. It appears as a line corresponding to the shortest path between the two turtles. A link is identified by the who number of the two turtles it links. Links can hold variables, and you can have different breeds of links.
\end{description}

Note that by default the NetLogo world uses toroidal boundary conditions meaning that anything which disappears off one edge of the world reappears on the opposite side. You can change this (in NetLogo 2D only) under `Settings...' by unchecking the `World wraps' checkboxes. For our two projects, we'll use the default periodic boundary conditions.

\subsection{Want to go further?}

The projects below show you what to do step-by-step. But if you want to extend these models or develop your own, the following resources will provide you with the information you need:
\begin{description}
\item[NetLogo User Manual] You will find it at \url{http://ccl.northwestern.edu/netlogo/docs}. In particular, under Reference, check out the Programming Guide and the NetLogo Dictionary (from the left-hand side menu).
\item[Models Library] You will find those within NetLogo under `File' $\rightarrow$ `Models Library'. In particular, check out the `Code Examples' models which are very helpful. Most of these codes are also available online at \url{http://ccl.northwestern.edu/netlogo/models}, but not the `Code Examples', unfortunately. Read the Models' description to see if it is likely to contain the type of code/behaviour you are looking to code-up in your own simulation.
\end{description}

\newpage

\section{Project: T cell movement within lymph nodes}

\begin{center}
\resizebox{0.7\linewidth}{!}{\includegraphics{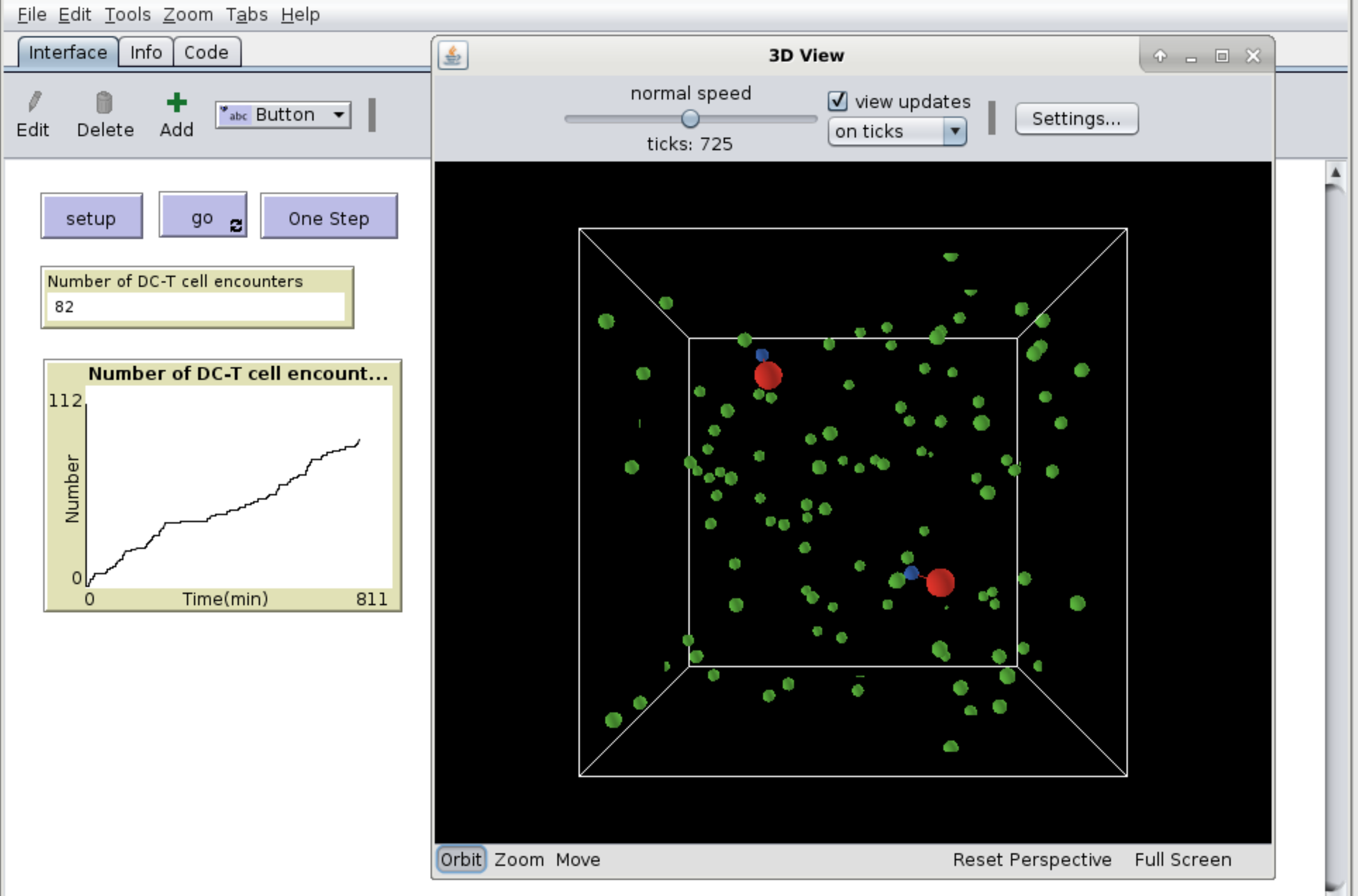}}
\end{center}

\subsection{Project description}

This project consists of T cells moving around within a 3D lymph node, encountering antigen-loaded dendritic cells (DCs) and making brief contacts. You will first create 100 T cells and have them move around randomly. You'll then introduce 2 DCs and when a T cell encounters a DC, the T cell will stop moving for \unit{3}{\minute} before resuming its regular motion. You will keep track of how many T cells have encountered a DC and you'll plot how that number evolves over time. You will also plot the tracks of some T cells, much like what gets recorded during two-photon microscopy experiments. Some extensions of this project include: making each DC bear different antigens and tracking how many cells have encountered each DC and how many have encountered both, tracking how the number of contacts changes for different T cell movement models.

\subsection{Step-by-step instructions}

\subsubsection{Creating your set of T cells and DCs}

This project is in 3D so you will need to use NetLogo 3D. Since your T cells and DCs will be moving around, they will be turtles. To make things clearer, you'll create two breeds of turtles: T cells and DCs. At the very top of your `Code' tab, you should enter:
\begin{verbatim}
breed [tcells tcell]    ; the T cells
breed [DCs DC]          ; the dendritic cells (DCs)
\end{verbatim}
When creating a new breed, the first word is the plural name of the breed and the second is its singular name. For example, if you were creating a breed of mice, you'd probably want \verb|breed [mice mouse]|. Note that `\verb|;|' is used to indicate a comment: everything after the \verb|;| on a given line is ignored by NetLogo.

You'll create the T cells and DCs as circles which map to spheres in NetLogo 3D. You'll make the DCs twice as large as the T cells. You'll create 100 green T cells and 2 red DCs. So let's write the `setup' procedure to do this:
\begin{verbatim}
to setup
  clear-all
  set-default-shape turtles "circle" ; all turtles will be circles (spheres)
  create-tcells 100 [
    set color green                  ; make T cell green
    setxyz random-xcor random-ycor random-zcor ; put T cell at random location
  ]
  create-DCs 2 [
    set size 2       ; make DC of size 2 (twice that of T cells)
    set color red    ; make DC red
    setxyz random-xcor random-ycor random-zcor
  ]
  reset-ticks
end
\end{verbatim}

Let's link the \verb|setup| procedure to a `setup' button. Go to your `Interface' tab. Make sure `Button' is selected, press `Add', and click anywhere in the `Interface' window. Under `Commands' type \verb|setup|, i.e.\ the name of the procedure you just created. Every time you click the `setup' button, your \verb|setup| procedure is run. Click it a few times to see how your initial setup changes.

\subsubsection{Setting the T cells in motion}

You'd like to make your T cells move. To begin with, you'll make them move randomly, but with a preference for continuing in their current direction (i.e.\ no drastic turn-arounds). You'll create a \verb|go| and a \verb|move-tcells| procedure just below your \verb|setup| procedure in your `Code' tab:
\begin{verbatim}
to go
  move-tcells
end

to move-tcells
  ask tcells [
    right random-normal 0 90 ; Pick random turn angle: avg = 0 std dev = 90 deg
    roll-right random-normal 0 90 ; Pick random roll angle
    forward 1
  ]
end
\end{verbatim}
The \verb|right 90| command would turn your turtle to the right by 90\degree. Here, instead of specifying the exact angle, we use \verb|random-normal 0 90| which will pick a random angle from a normal distribution of mean zero and standard deviation 90\degree.

Now, create a button as before, but this time enter `go' in the `Command' and check the `Forever' checkbox. Click the button: wow, look at these T cells go! You can adjust the speed with the slider.

But perhaps you'd like to create a button to see the T cells move by one step only. Sure, create a button as before, enter `go' in the `Command', but this time, do not check the `Forever' checkbox and you might want to change the display name to `one step' or something similar to distinguish it from the other `go' button.

\subsubsection{Let's get Physical}

Alright, we have nice little balls moving around a grid, travelling at a speed of one patch per time step. What is that?! We need to map the time and space in our model to REAL physical time and space. Here is how to do this.

Since we did not specify the size of our \verb|tcells|, NetLogo created them with a diameter of one (1). So without even noticing, we have decided that a distance of 1 in our simulation space corresponds to a physical distance of \unit{8}{\micro\metre} in real life, i.e.\ the diameter of a T cell. This seems like a good choice so we'll leave it like that. As for time, let's decide that each time step of our simulation corresponds to \unit{0.5}{\minute} (or \unit{30}{\second}). So let's define
\bea
\text{dt} &=& \frac{\unit{0.5}{min}}{\unit{1}{time\ step}} \\
\text{ds} &=& \frac{\unit{8}{\micro\metre}}{\unit{1}{patch}}
\eea

Based on the two-photon microscopy literature, we know T cells move at a speed of about \unit{16}{\micro\metre/\minute}. How do we convert that to units of patch per time step for use in NetLogo?
\bea
v_\text{NetLogo} &=& \left(\frac{\unit{16}{\micro\metre}}{\unit{1}{\minute}} \times \frac{\unit{0.5}{\minute}}{\unit{1}{time\ step}}\right) \times \frac{\unit{1}{patch}}{\unit{8}{\micro\metre}} \\
&=& \left(\frac{\unit{8}{\micro\metre}}{\unit{1}{time\ step}}\right) \times \frac{\unit{1}{patch}}{\unit{8}{\micro\metre}} \\
&=& \frac{\unit{1}{patch}}{\unit{1}{time\ step}}
\eea
So, more generally, if you want to convert your real speed to a NetLogo speed, you'd write:
\be
v_\text{NetLogo} = v_\text{real life} \times \frac{\text{real time}}{\unit{1}{time\ step}} \times \frac{\unit{1}{patch}}{\text{real space}} = v_\text{real life} \times \frac{\text{dt}}{\text{ds}}
\ee

So let's adjust our simulation accordingly. At the top of your file, above all procedures, add the following line
\begin{verbatim}
globals [
  dt  ; dt = duration of one step in min/time step
  ds  ; ds = size of one patch in um/patch
  tcell-step ; tcell-step = size of tcell step in patches/time step
]
\end{verbatim}
Then in your \verb|setup| procedure, just above \verb|reset-ticks|, add
\begin{verbatim}
set dt 0.5  ; 0.5 min per time step
set ds 8    ; 8 um per patch
set tcell-step 16 / ds * dt ; (16 um/min)*(1 patch/8 um)*(0.5 min/time step)
\end{verbatim}
and modify your \verb|go| and \verb|move-tcells| procedures to be
\begin{verbatim}
to go
  move-tcells
  tick-advance dt
end

to move-tcells ; procedure for T cell motion
  ask tcells [
    right random-normal 0 90
    roll-right random-normal 0 90
    forward tcell-step
  ]
end
\end{verbatim}
Now your tick counter in the 3D View is in minutes. Let's adjust its label accordingly. Click on `Settings' and under `Tick counter label' enter `Time (min)', and hit `Ok'. I also recommend you set the update to `on ticks' rather than `continuous'.

\subsubsection{Designing T cell-DC contacts}

Two-photon microscopy has taught us that when a T cell encounters a DC, it will pause for about \unit{3}{\minute} before resuming its regular motion. We also know that DCs have dendrites approximately \unit{19}{\micro\metre} \cite{miller04pnas}. Therefore, we can design T cell-DC contact by imposing the following two rules:
\begin{itemize}
\item A T cell is considered in contact with a DC if it is within a radius of \unit{19}{\micro\metre} of the DC. Note that for us, this corresponds to a distance of 2 patches ($\unit{19}{\micro\metre} \times \unit{1}{patch}/(\unit{8}{\micro\metre})$).
\item A T cell stops moving for \unit{3}{\minute} once it is in contact with a DC.
\end{itemize}
For this purpose it might make sense to use links. We will create a link between a T cell and a DC when the distance requirement is met, and we could give the link a property \verb|tDCfree| which would keep track of the time left before the T cell is free from the DC. 

At the top of the file, before the procedures, add
\begin{verbatim}
links-own [ tDCfree ]
\end{verbatim}
This states that all \verb|links| will now have a variable called \verb|tDCfree|. It is like \verb|globals| variables, but it belongs only to a specific type of agent, in this case the \verb|link| agents.

What we want is to:
\begin{itemize}
\item Create a directed link from the DC to all T cells within a 2 patch diameter.
\item Make the link red, just like the DC.
\item Set the \verb|tDCfree| to \unit{3}{\minute}
\item Make the linked T cells blue
\end{itemize}
Here is how you do this in NetLogo. First, add a new procedure we'll call \verb|identify-DCbound-tcells| in \verb|go|
\begin{verbatim}
to go
  move-tcells
  identify-DCbound-tcells
  tick-advance dt
end

to identify-DCbound-tcells ; procedure to identify DC-bound T cells
  ask DCs [
    create-links-to tcells with [ color != blue ] in-radius 2 [
      set color red  ; make the link red
      set tDCfree 3  ; bound for 3 minutes
      ask end2 [ set color blue ] ; make the linked T cell blue
    ]
  ]
end
\end{verbatim}
For each DC, \verb|create-links-to| creates a link from that DC to all T cells whose colour is not blue which are located within a radius of 2 patches. The \verb|with [color != blue]| ensures that only T cells that are not blue (i.e.\ that are not already linked to a DC) are considered.

The \verb|ask end2 [...]| means ask the turtle at the end of the link to become blue. In this case, \verb|end2| is a T cell since the directed link is from the DC to the T cell.

It's good to make your changes incrementally. Now check that your code still works as before by pressing the `setup' and `go' buttons. If all is still working, you're ready to go to the next step.

That's great, but now the T cells remain blue forever and their link to the DCs are never removed. First, add a new procedure we'll call \verb|update-link-status| to the \verb|go| procedure, right below \verb|identify-DCbound-tcells|. Second, add the actual procedure at the bottom of your code as follows:
\begin{verbatim}
to update-link-status ; function to update the DC-T cell links
  ask links [
    set tDCfree tDCfree - dt  ; update time remaining for link
    if tDCfree < 0 [  ; if tDCfree has elapsed
      ask end2 [ set color green ] ; set T cell to its green colour
      die ; kill the link (delete it)
    ]
  ]
end
\end{verbatim}
Wow, this is awesome! But I sure wish we could track of how many encounters took place\ldots

\subsubsection{Counting the DC-T cell encounters}

It would be nice to know how many DC-T cell encounters occurred. That's where a \verb|Monitor| comes in handy. First, create a new global variable to keep track of DC-T cell encounters: we'll call it \verb|nDCmeet|. So
\begin{verbatim}
globals [
; ... the other stuff you already have here
  nDCmeet ; number of DC-T cell encounters
]
\end{verbatim}
and then somewhere within \verb|create-links-to| in the \verb|identify-DCbound-tcells| procedure, you can add
\begin{verbatim}
set nDCmeet nDCmeet + 1
\end{verbatim}
Now, in your `Interface' tab, add a `Monitor'. Under `Reporter' enter \verb|nDCmeet| and under `Display name' something like \verb|Number of DC-T cell encounters| for clarity. Now run the simulation and see how the Monitor gets updated and keeps growing as more and more encounters occur.

\subsubsection{Plotting the DC-T cell encounters}

Ah, yes, a graph would be nice indeed. Here's how you do this. In your \verb|to go| function, just above \verb|tick-advance dt| add the following line
\begin{verbatim}
plotxy ticks nDCmeet
\end{verbatim}
This will plot the number of DC-T cell contacts made thus far (\verb|nDCmeet|) as a function of time (\verb|ticks|).

In your `Interface' tab, add a `Plot', call it something meaningful like \texttt{Number of DC-T cell encounters} and set appropriate names for the $x$ and $y$ axis labels (e.g., \texttt{time (min)}, and \texttt{Number}). Now run the simulation and see what happens... cool!

\subsection{Open-ended problems}

Well, that was fun. Now it's your turn! Below are a few challenging additions for you to work into your code. If you have time, dive in and see how far you can get. Remember that you can consult the \href{http://ccl.northwestern.edu/netlogo/docs}{NetLogo User Manual} or the \href{http://ccl.northwestern.edu/netlogo/models}{Models Library} for additional information.

\subsubsection{A 2-antigen system}

This extension consists of making each of the 2 DCs bear different antigens and tracking how many cells have encountered each DC, and how many have encountered both. You could, for example, assign a different colour to your 2 DCs and have T cells change their colour to match that of the DC they have encountered, and have each T cell record internally how many times they have encountered each DC. You could also use colour to visually identify those T cells which have encountered both.

\subsubsection{The effect of T cell movement}

This extension consists of exploring how different modes of movement for T cells affect the number of DC-T cell encounters over the course of the simulation. For inspiration for T cell movement, check out these references: \cite{beltman07,mempel04,miller02}.

\subsubsection{Random walk versus chemotaxis}

This extension consists of exploring the effect of chemokine-biased T cell movement. You could make your DCs secrete chemokines (into their patch), have the chemokines diffuse (check out the \texttt{diffuse} function) to neighbouring sites, and have T cells move preferentially towards the sites with more chemokines (check out the \texttt{uphill} function).

\newpage

\section{Project: Viral infection spread in vitro}

\begin{center}
\resizebox{0.8\linewidth}{!}{\includegraphics{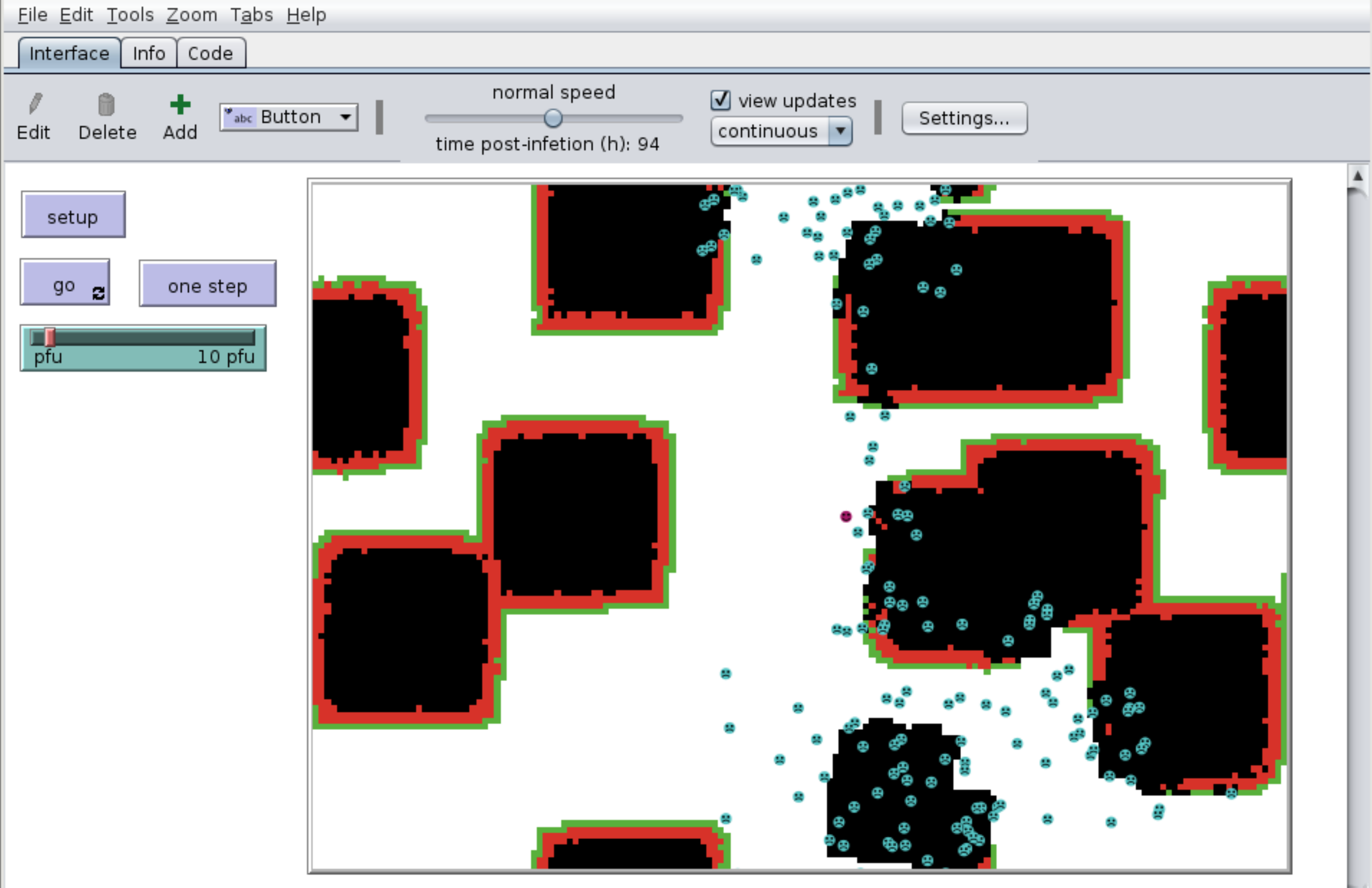}}
\end{center}

\subsection{Project description}

This project aims to model the progress of an influenza viral infection through an in vitro cell culture. The simulation grid will represent your confluent cell culture with each site of the square grid (patch) representing one cell. As cells become infected, they will undergo a change of state (colour) going from uninfected (white) to latently infected (green) to infectious (red) to dead (black). At each time step, you will determine if uninfected cells become infected based on the number of infectious neighbours they have, and if latently infected cells become infectious or infectious cells die based on how much time has elapsed since these cells were infected. You will also add the action of cytotoxic T lymphocytes moving around the grid randomly killing infectious cells. Some extensions of this project include: plotting the fraction of cells in each state over time to monitor the progression of the infection, adding the action of interferon or antivirals, or modelling extracellular virus explicitly, or enabling cells to divide to repopulate the holes left behind by dead cells.

\subsection{Step-by-step instructions}

\subsubsection{Setting up our cell culture}

Since this project involves static cells, we will use patches to represent each cell and colour to represent their state. We'll want to initialize our infection with all cells uninfected (white), except for a few of them in the latently infected state. For now, we'll make 5 initially latently infected cells.

In your `Code' tab, you should enter
\begin{verbatim}
to setup
  clear-all
  ; setup whole grid
  ask patches [
    set pcolor white ; uninfected
  ]
  ; setup initially infected cells
  ask n-of 5 patches [
    set pcolor green ; latently infected
  ]
end
\end{verbatim}

Note that `;' is used to indicate a comment: everything after the `;' is ignored by NetLogo so it allows you to leave yourself notes in the code to remember what the commands do. The \verb|n-of 5| command tells NetLogo to pick 5 patches at random.

Now, let's link the \verb|setup| function to a `setup' button. Go to your Interface tab. Make sure `Button' is selected, press `Add', and click anywhere in the Interface window. Under `Commands' enter \verb|setup|, i.e.\ the name of the function you just created. Every time you click the `setup' button, your \verb|setup| procedure is run. Click it a few times to see how your initial setup changes.

\subsubsection{Controlling the number of pfu in our inoculum}

Instead of having 5 cells infected, we want to be able to set the number of cells to be initially infected (i.e., the number of plaque forming units or pfu in the inoculum) directly from the `Interface' tab. Let us call that number \verb|pfu|.

First, add a `Slider' to the `Interface' tab, following the same procedure you used for adding a button above, and enter the following information in the entry fields:
\begin{center}
\begin{tabular}{lc}
`Global variable' & \verb|pfu| \\
`Minimum' & 1 \\
`Maximum' & 100 \\
`Value' & 5 \\
`Units (optional)' & pfu
\end{tabular}
\end{center}
leaving the rest of the fields as they were.

Second, replace the \verb|n-of 5| with \verb|n-of pfu| in your `Code' tab. Now if you go back to your `Interface' tab, you can try moving the slider to different \verb|pfu| values and press `setup' and verify that your slider successfully changes the number of initially infected cells. Ah, look at you: now you control the initial inoculum. Such power!

\subsubsection{Deciding who lives or die}

Ok, so we have infected cells: now what? Well, these newly infected or eclipse cells, after some time, will begin synthesizing viral proteins and soon they should be able to release virus, becoming infectious. And after some time producing virus, these infectious cells will cease viral production and undergo apoptosis either because they have exhausted the available nutrients making all these virions, or due to toxicity resulting from viral production.

To reproduce this behaviour, we'll have each cell keep track of when it should become infectious and when it should die based on when it was infected. In your `Code' tab, \underline{above} \verb|to setup|, we will define the following patch properties:
\begin{verbatim}
patches-own [tinfectious tdead]
\end{verbatim}

Now we have to make sure that, when we initially infect our \verb|pfu| cells, all of them decide right then when they should become infectious and die (i.e., set the value of their \verb|tinfectious| and \verb|tdead|. So \underline{inside} \verb|to setup|, replace your \verb|ask n-of pfu| section with this one:
\begin{verbatim}
ask n-of pfu patches [
  set pcolor green
  set tinfectious 4
  set tdead tinfectious + 7
]
reset-ticks
\end{verbatim}
so now our \verb|pfu| infected cells have decided that they will become infectious after 4 ticks and will die after 11 ticks (4 + 7). NetLogo uses `ticks' to keep track of time.

Now we need to make time advance and keep track of whether these cells are ready to change state. Below the \verb|end| command marking the end of the \verb|setup| procedure, add a new \verb|go| procedure.
\begin{verbatim}
to go
  ; check if latently infected cells become infectious
  ask patches with [pcolor = green] [
    if tinfectious <= ticks [
      set pcolor red
    ]
  ]
  ; check if infectious cells die
  ask patches with [pcolor = red] [ 
    if tdead <= ticks [
      set pcolor black
    ]
  ]
  tick-advance 1 ; advance time by one time step
end
\end{verbatim}

To run this procedure, create a button as before, but this time enter `go' in the `Command'. Once the `go' button has been created, click the `setup' button and then the new `go' button: the `ticks:' number at the top of your simulation grid go up every time you click on `go' and when you get to `ticks: 5', all your cells become red (infectious). If you keep clicking, they'll turn black (dead).

But all this clicking is not good for your wrist. It would be nice if it could just keep going. Right-click on the `go' button, select `Edit...' and check the `Forever' checkbox. Now, if you click the `go' button, it will go so fast that you won't see a thing. You can adjust the speed at which things happen by moving the slider at the top of the `Interface' tab which defaults to `normal speed' but can be adjusted as you wish.

If you would like to keep the option of being able to make the simulation advance by just one step, create another button with the command \verb|go|, but do not check `Forever' and set the `Display name' to something like `one step' to distinguish it from the `go' button. Now you can `go' continuously or one step at a time.

\subsubsection{Making it big}

It would be nice to look at a bigger patch of tissue. By default, NetLogo creates a grid that is $33\times33$ where each square patch is 13 pixels wide. Here, we would prefer to have a higher resolution image with more cells to look at.

In your `Interface' tab, click on `Settings...' and set:
\begin{center}
\begin{tabular}{lc}
\verb|max-pxcor| & 80 \\
\verb|max-pycor| & 56 \\
Patch size & 6
\end{tabular}
\end{center}
Oh, much better. Click the setup and go button to see what things look like now.

\subsubsection{Let's get Physical}

What we have now is patches which evolve as time in units of ticks goes by... we need to connect time and space to real-life dimensions and units for our results to ultimately be meaningful. Space-wise, let us say that one patch = one cell: that is simplest. Time-wise, let us use units of hours. Now we need to reflect that choice in our code.

Each time step currently lasts 1 tick, and upon infection it will take your cells 4 ticks to begin producing virus, and 11 ticks to die. In reality, for an influenza infection, the eclipse phase, i.e.\ the pause between a cell's successful infection by a virion and release of the first infectious virion, takes $\sim$\unit{6}{\hour} \cite{holder11h1n1}. An infectious cell will produce virus for $\sim$\unit{12}{\hour} before ceasing viral production and undergoing apoptosis. Note that these times are approximate and vary for different cell cultures, viral strains, and experimental conditions.

Let us define three global variable to help us keep time of the duration of a time step, the eclipse phase, and the infectious phase. At the top of our `Code', above \verb|patches-own| add the following lines
\begin{verbatim}
globals [
  dt ; time step duration (h)
  eclipsedur ; duration of eclipse phase (h)
  infectiousdur ; duration of infectious phase (h)
]
\end{verbatim}
It is always good to leave yourself comments about what your variables do and what units they are in. We now need to set the value of these 3 variables. Inside \verb|to setup|, right after \verb|clear-all| add
\begin{verbatim}
  set dt 0.1              ; 0.1h (6 min)
  set eclipsedur 6.0      ; 6h
  set infectiousdur 12.0  ; 12h
\end{verbatim}
Still within setup, in the \verb|n-of pfu patches| section, replace
\begin{verbatim}
  set tinfectious 4
  set tdead tinfectious + 7
\end{verbatim}
with
\begin{verbatim}
  set tinfectious eclipsedur
  set tdead tinfectious + infectiousdur
\end{verbatim}
To impose the duration of a time step, \verb|dt|, in \verb|to go|, replace the command \verb|tick-advance 1| with \verb|tick-advance dt|. And to make these time-units obvious when looking at the simulation, go to your `Interface' tab, click on `Settings...' and under `Tick counter label' enter something like `time post-infection (h)', and click `Ok'.

\subsubsection{Spreading the joy... of infection}

You might have noticed that right now your initially infected cells, though they become red and are supposedly infectious, do not infect other cells. This is what we will fix now. We could get complicated and allow cells to release virus and let virus diffuse outwards and infect other cells... but that it a little too ambitious (see Section Open-ended problems below).

For now, let us assume infection is taking place under a thick agar so that infectious cells can only infect their immediate neighbours. And let us assume that an infectious cell can infect one of its uninfected neighbours every \unit{6}{min} (that's \unit{0.1}{\hour} or one time step in our simulation) until there are none left.

In `Code', immediately after \verb|to go|, add the following line
\begin{verbatim}
to go
  spread-the-infection
\end{verbatim}
and after the entire \verb|to go| procedure, below \verb|end|, add
\begin{verbatim}
to spread-the-infection
  ask patches with [pcolor = red] [
    let nnei count neighbors with [pcolor = white]
    if nnei > 0 [ ; if more than 0 neighbours are uninfected
      ask one-of neighbors with [pcolor = white] [ ; ask one of them
        set pcolor green
        set tinfectious ticks + random-normal eclipsedur (0.1 * eclipsedur)
        set tdead tinfectious + random-normal infectiousdur (0.1 * infectiousdur)
      ]
    ]
  ]
end
\end{verbatim}
The command \verb|let nnei count neighbors with [pcolor = white]| will set variable \verb|nnei| equal to the number of uninfected neighbour the infectious cell has. And if it has more than zero, it proceeds to infect one of them.

The command \verb|random-normal eclipsedur (0.1 * eclipsedur)| means that the duration of the eclipse phase for each newly infected cell will be chosen at random from a normal distribution of mean \verb|eclipsedur| and a standard deviation of 10\% of \verb|eclipsedur|. Having these durations be random is more realistic because not all cells will become infected at the same time. The same is true for \verb|tdead|.

Time to test out your infection spread. Go back to `Interface' and type `go'. Cool!

\subsubsection{Sending in the death squad: adding cytotoxic T lymphocytes}

This might not be entirely realistic for an in vitro cell culture, but it is a fun addition and will allow you to see how to manipulate moving agents (turtles) since so far we have focused exclusively on patches.

We will originally have 2 naive CD8+ T cells roaming the grid, moving randomly from site-to-site. Upon encounter of an infectious cells, the naive CD8+ T cell becomes an effector CTL. When an effector CTL encounters an infectious cell, it kills it. Effector CTLs divide with a period of \verb|divtime| to a maximum of \verb|maxdiv|, after which it dies.

We will have to create a new breed of turtles: T cells. At the top of your `Code' file, write
\begin{verbatim}
breed [tcells tcell]
tcells-own [ ttdead tdiv ndiv ]
\end{verbatim}
and add the following variables to \verb|globals|
\begin{verbatim}
globals [
  divtime ; time between CTL division (h)
  maxdiv  ; max number of CTL divisions
\end{verbatim}
In \verb|to setup| you will need to add the following lines
\begin{verbatim}
  set divtime 7.0
  set maxdiv 7
  set-default-shape turtles "face happy"
  create-tcells 2 [
    set size 2
    set color magenta
    setxy random-xcor random-ycor
  ]
\end{verbatim}
which sets the time between T cells divisions to \unit{7}{\hour}, and the number of divisions the CTL will undergo before contraction to 7 division. Additionally, it specifies that the default shape for T cells will be a \verb|happy face| symbol. It then adds 2 magenta (naive) T cells at random locations on the grid.

As for letting these cells move about, mature, divide and die, you will need to add the line \verb|update-tcells| just above \verb|tick-advance dt| in \verb|to go|, and at the bottom of the `Code' file, add the following new procedure:
\begin{verbatim}
to update-tcells
  ask tcells [
    let pclr [pcolor] of patch-here
    if (pclr = green) or (pclr = red) [
      ifelse color = magenta [
        set color cyan
        set shape "face sad"
        set tdiv ticks + random-normal divtime 1.0
        set ndiv 0
      ][
        ask patch-here [ set pcolor black ]
        if tdiv <= ticks [
          ask patch-here [
            sprout-tcells 1 [
              set shape "face sad"
              set color cyan
              set size 2
              set tdiv [tdiv] of one-of other tcells-here
              set ndiv [ndiv] of one-of other tcells-here
            ]
          ]
          set tdiv ticks + random-normal divtime 1.0
          set ndiv ndiv + 1
          if ndiv > maxdiv [
            die
          ]
        ]
      ]
    ]
    right random 360
    forward 1
  ]
end
\end{verbatim}

\subsection{Open-ended problems}

Below are a few challenging additions for you to work into your code. If you have time, dive in and see how far you can get. Remember that you can consult the \href{http://ccl.northwestern.edu/netlogo/docs}{NetLogo User Manual} or the \href{http://ccl.northwestern.edu/netlogo/models}{Models Library} for additional information.

\subsubsection{Plotting the number of cells in each state}

Using the command \verb|plotxy|, you can plot the number of cells in each state over time. This will allow you to monitor the progression of the infection, much like one could measure using FACS (at least for infected cells).

\subsubsection{Adding the action of antivirals}

To simulate the action of antivirals, you could have a button in your `Interface' tab which turns the antiviral on at any time you wish. Its effect could be, for example, to lengthen the time it takes for cells to begin producing virus (\verb|eclipsedur|).

\subsubsection{Implementing cell division}

It would be nice if uninfected cells could divide to replenish the dead cells. For example, if cells divide rapidly to regenerate dead cells, it might allow the infection to move back into an area it had destroyed and this in turn could lead to chronic infection. A sensible rule would be to give uninfected cells a probability to turn one of its dead (black) neighbour into an uninfected (white) neighbour at every time step.

\vspace{1em}
\hrule
\vspace{1em}

\addcontentsline{toc}{section}{References}
\bibliographystyle{abbrvurl}
\bibliography{allbibliographies}

\end{document}